\documentstyle[12pt,epsf]{article}
 \font\blackboard=msbm10 
 \font\blackboards=msbm7 \font\blackboardss=msbm5
 \newfam\black \textfont\black=\blackboard
 \scriptfont\black=\blackboards \scriptscriptfont\black=\blackboardss
 \def\Bbb#1{{\fam\black\relax#1}}

\font\ninerm=cmr9
\def\uniset{\rlap{\ninerm 1}\kern.15em 1}

\def\d{{\rm d}}
\def\mi{{\rm i}}
\def\j{{\rm j}}
\def\G{\mathop{\Gamma}\nolimits}
\def\Re{\mathop{\rm Re}\nolimits}
\def\e{\mathop{\rm e}\nolimits}
\def\sq2{\sqrt{2}}

\def\defi{\stackrel{\rm def}{=}}
\newcommand\grsim{\mathrel{\hbox{\lower1ex\hbox{\rlap{$\sim$}\raise1ex\hbox{$>$}}}}}
\newcommand\losim{\mathrel{\hbox{\lower1ex\hbox{\rlap{$\sim$}\raise1ex\hbox{$<$}}}}}

\title{An exact solution method for 1D polynomial Schr\"odinger equations}
\author{{\bf Andr\'e Voros} \\
\\
CEA--Saclay, Service de Physique Th\'eorique\\
F-91191 Gif-sur-Yvette CEDEX (France)\\
{ E-mail : voros@spht.saclay.cea.fr}\\
\\
and\\
Institut de Math\'ematiques de Jussieu\\
CNRS UMR 7586\\
Universit\'e Paris 7\\
2 place Jussieu,
F-75251 Paris CEDEX 05 (France)}

\begin{document}
\maketitle
{\abstract 
Stationary 1D Schr\"odinger equations with polynomial potentials are reduced
to explicit countable closed systems of exact quantization conditions,
which are selfconsistent constraints upon the zeros
of zeta-regularized spectral determinants, complementing the usual asymptotic
(Bohr--Sommerfeld) constraints. (This reduction is currently completed under
a certain vanishing condition.) 
In particular, the symmetric quartic oscillators are admissible systems, 
and the formalism is tested upon them.  
Enforcing the exact and asymptotic constraints by suitable iterative schemes,
we numerically observe geometric convergence to the correct eigenvalues/functions in some test cases, suggesting that
the output of the reduction should define a contractive fixed-point problem
(at least in some vicinity of the pure $q^4$ case).
}
 
\section{The Schr\"odinger spectral problem}

We study the 1-dimensional stationary Schr\"odinger equation 
with a real polynomial potential $V(q)$ on the real axis
(\cite{BW}--\cite{V4} provide some directly related references).

In reduced form,
\begin{equation}
\label{Schr}
-\psi''(q) + (V(q) + \lambda) \psi(q) = 0, \qquad  
V(q) \equiv +q^N + v_1 q^{N-1} + v_2 q^{N-2} + \cdots + v_{N-1} q.
\end{equation}
We write the collection of coefficients as $\vec v \defi (v_1,\cdots,v_{N-1})$,
absorbing any constant term into the spectral parameter $\lambda$.

The case $N=2$ (singular) is excluded \cite{V4}.

\subsection{Polarized boundary conditions}

We first cast eq.(\ref{Schr}) into an eigenvalue problem 
with asymmetric boundary conditions. 
I.e., we restrict the problem to a half-line on which $V$ is confining:
$[0,+\infty)$, keeping the square-integrability at $+\infty$, 
then put a Neumann, 
resp. Dirichlet, boundary condition at the finite endpoint $q=0$.
Both conditions then define self-adjoint operators $\hat H^+$, resp. $\hat H^-$,
which have purely discrete spectra
respectively denoted $\{E_{2k}\}$ and $\{E_{2k+1}\}$ for $k=0,1,2, \ldots$; 
these admit an asymptotic expansion which is the Bohr--Sommerfeld formula
reexpanded in energy,
\begin{equation}
\label{BS}
\sum_{j=0,1,2,\ldots} {\tilde b}_{\mu-j/N} E_k^{\mu-j/N}
\sim 2\pi (k+1/2),\qquad k \to \infty {\rm \ in\ }\Bbb N, \quad {\rm with} 
\end{equation}
\begin{equation}
\label{MU}
\mu \defi {1 \over 2}+ {1 \over N}\ :\ {\rm the\ }growth\ order;\quad 
{\tilde b}_\mu \defi \oint_{p^2+q^N=1} p \d q =
{2\pi^{1/2}\over N}\G\Bigl({1 \over N}\Bigr) \Big/ \G\Bigl({3 \over 2}+{1 \over N}\Bigr) ;
\end{equation}
the higher ${\tilde b}_{\mu-j/N}$ are polynomials in the $v_{j'}\ (j' \le j)$,
e.g., ${\tilde b}_{\mu-1/N} = -4v_1/N$. Crucially, $\mu<1$.

Up to sign, the eigenvalues read as the zeros of the Fredholm determinants
\begin{equation}
\label{Fred}
\Delta^\pm (\lambda) \defi 
\prod_{k\ {\scriptstyle{\rm even}\atop \scriptstyle{\rm odd}}}
(1+ \lambda / E_k);
\end{equation}
these infinite products converge to entire functions of order $\mu$ in the variable $\lambda$, also entire in the parameters $\vec v$ \cite{S}.

At fixed $(\lambda,\vec v)$, 
let $\psi_\lambda (q) $ denote a recessive solution of eq.(\ref{Schr}) 
(i.e., a solution exponentially decreasing as $q \to +\infty$, 
which is unique up to normalization).
Then $ \psi_\lambda (0) $ vanishes simultaneously with $ \Delta^- (\lambda)$,
(and likewise for $ \psi'_\lambda (0) $ and $ \Delta^+ (\lambda) $).
However, the relations connecting them retain messy factors unless 
each of $\Delta $ and $ \psi$ is suitably renormalized.
The procedure for the determinants is the well-known zeta-regularization,
but we will also introduce a parallel operation upon the wave functions.

\subsection{The spectral (or functional) determinants}

By Mellin-transforming eq.(\ref{BS}), the spectral zeta functions, \cite{V5}
\begin{equation}
\label{Z} 
Z^\pm (s) \defi 
\sum_{k\ {\scriptstyle{\rm even}\atop \scriptstyle{\rm odd}}} \lambda_k^{-s}
\qquad  ({\rm and} \quad Z(s) \defi Z^+ (s) +Z^- (s) ),
\qquad  \qquad (\Re s > \mu)
\end{equation}
are shown to extend {\it meromorphically\/} to all of $\Bbb C$, 
and to be regular at integer points.

In particular, the point $s=0$ (where $Z(0)=(2\pi)^{-1}{\tilde b}_0$)
is used to define determinants as {\sl zeta-regularized products\/},
\begin{equation}
\label{Det}
\det \hat H^\pm \defi \exp [- {Z^\pm}'(0)] \quad {\rm and\ likewise,}
\quad D^\pm(\lambda) \defi \det (\hat H^\pm + \lambda).
\end{equation}
This normalization, unlike that in eq.(\ref{Fred}), commutes with 
global spectral translations.

More explicit formulae are
\begin{equation}
D^\pm(\lambda) \equiv \exp [- {Z^\pm}'(0)]\, \Delta^\pm (\lambda);
\end{equation}
\begin{eqnarray}
\label{McL}
\log D^\pm(\lambda)  &=& \lim _{K \to +\infty} \Bigl\{ \sum_ {k<K}
\log (E_{k} + \lambda) + {1\over 2} \log (E_{K} + \lambda) \\
&&\qquad -{1 \over 4\pi}\sum_{0 \le j < \mu N} {\tilde b}_{\mu-j/N} E_{K}^{\mu-j/N}
\Bigl( \log E_{K}-{1 \over \mu-j/N}\Bigr)
 \Bigr\} \quad
{\rm for\ }k,K\ \scriptstyle{\scriptstyle{\rm even}\atop \scriptstyle{\rm odd}}.
\nonumber
\end{eqnarray}

They will also be used for complex spectra, as justified in \cite{QHS}.

\subsection{Natural WKB normalization for recessive solutions}

A recessive solution of eq.(\ref{Schr}) admits an exact WKB representation,
\begin{equation}
\label{WKB}
 \psi_{\lambda,q_0} (q) \equiv u(q)^{-1/2} \e^{-\int_{q_0}^q u(q') \d q'}
\quad {\rm where,\ for\ } q \to +\infty , \quad
u(q) \sim
\Pi (q) \defi (V(q) + \lambda)^{1/2}.
\end{equation}
This normalization awkwardly depends on the base point $q_0$, 
so we intend to define it for $q_0=+\infty$:
\begin{equation}
\label{NN}
\psi_\lambda (q) \defi u(q)^{-1/2} \e^{\int_q^{+\infty} u(q') \d q'} \defi 
\e^{\int_0^{+\infty} u(q) \d q} \psi_{\lambda,0}(q).
\end{equation}
However, only $u-\Pi$ is integrable at infinity whereas $u \sim q^{N/2}$, 
so a further prescription is needed, i.e., 
\begin{equation}
\label{zz}
\int_0^{+\infty} u \d q \defi 
\int_0^{+\infty} (u - \Pi) \d q + {\mathcal I}(s,\lambda)|_{s=-1/2},\qquad
{\mathcal I}(s,\lambda) \defi \int_0^{+\infty} (V(q) + \lambda)^{-s} \d q ,
\end{equation}
the latter meaning the analytical extension of the formula 
from its convergence domain $\{\Re s > 1/N,\ \lambda > -\inf V \}$).
As an example,
\begin{equation}
\int_0^{+\infty} \Pi(q) \d q = {{\tilde b}_\mu \over 4\sin\pi\mu} \lambda ^\mu
\qquad {\rm for\ } V(q) \equiv q^N. 
\end{equation}

In general, however, ${\mathcal I}_s$ develops a pole at $s=-1/2$ with residue
$-Z(0)/2$, requiring further regularization in general. 
All present calculations are carried out under the simplifying assumption 
$Z(0) \equiv 0$, a special case which still includes whole classes
of potentials unconditionally, for instance: $V$ homogeneous, or purely odd, 
or purely even with $N$ multiple of 4.

The new normalization (\ref{NN}) is then well defined, 
and it commutes with spatial translations.

\subsection{Basic identities}

Under the above notations and restrictions, very simple identities connect
the spectral determinants and the naturally normalized solution:
\begin{equation}
D^-(\lambda ) \equiv \psi_\lambda (0), \qquad 
D^+(\lambda ) \equiv -\psi'_\lambda (0).
\end{equation}
(The proof is an adaptation of the arguments in \cite{V1}, Apps.~A and D).

Next, following \cite{S}, we continue eq.(\ref{Schr}) in the complex $q$-plane
down to the rotated half-line having the adjacent Stokes direction, namely
$ [0,\e^{-\mi\varphi/2 }\infty) $ with
\begin{equation}
\label{ID}
\varphi \defi {4 \pi \over N+2} \quad  (spectral\ symmetry\ angle).
\end{equation}
By simple complex scaling upon $q$,
\begin{equation}
\psi^{[1]} \defi \psi_{ \e^{-\mi\varphi} \lambda} 
(\e^{\mi\varphi/2}q ; {\vec v}^{[1]})
\end{equation}
provides a recessive solution to eq.(\ref{Schr}) 
in that Stokes direction, where
\begin{equation}
{\vec v}^{[1]} \defi (\e^{\mi\varphi/2} v_1, \e^{2\mi\varphi/2} v_2, \cdots,
\e^{(N-1)\mi\varphi/2} v_{N-1})
\end{equation}
expresses an action of the discrete rotation group of order $(N+2)$ on the coefficients; equivalently it acts upon the potential $V$, mapping it to
$V^{[1]}$, then $V^{[2]}, \cdots$ (now {\sl complex\/} potentials). 
We call $L$ the order of the effective symmetry group 
($L=N+2$ generically, while $L={N \over 2}+1$ for an even $V$).

Now, the Wronskian of the two solutions $\psi, \psi^{[1]}$ of eq.(\ref{Schr}) is
a constant, which can be evaluated as $q \to \infty$ because the domains of
validity of their respective asymptotic expansions (\ref{WKB}) overlap,
and also expressed at $q=0$ by means of the respective identities (\ref{ID})
for the potentials $V$ and $V^{[1]}$. 
Matching the two calculations then yields the fundamental bilinear identity
\begin{equation}
\label{DW}
\e^{+\mi\varphi/4} D^+( \e^{-\mi\varphi} \lambda, {\vec v}^{[1]} )
D^-( \lambda, {\vec v})
-\e^{-\mi\varphi/4} D^+( \lambda, {\vec v}) 
D^-( \e^{-\mi\varphi} \lambda, {\vec v}^{[1]} ) \equiv 2 \mi.
\end{equation}

We stress that our approach skips any other matching of solutions,
such as those in connection problems between nonadjacent Stokes directions,
which yield nontrivial Stokes multipliers...

We also mention that similar identities have appeared in quantum integrable
systems, as ``quantum Wronskians" \cite{QW}.

\section{Exact quantization conditions}

For a homogeneous potential, we currently believe that the functional relation
(\ref{DW}) and the asymptotic law (\ref{BS}) (imposed to some o(1) accuracy) 
together suffice to specify the whole spectrum exactly:
indeed, we can recover the spectrum within each parity sector as the fixed
point of an (apparently) contractive mapping built upon those data alone
\cite{V2}\cite{V3}\cite{V4}. 
Now we start to extend this approach towards more general polynomial potentials,
still keeping $Z(0)=0$ (a broader generalization is under study).

Guided by the homogeneous case \cite{V3}, and in order to obtain
equations for the eigenvalues of a given parity, say even,
(the odd case admits a symmetrical separate treatment), we set $\lambda=-E_{2k}$
in eq.(\ref{DW}) and its partner with ${\vec v}^{[-1]}$, 
then eliminate the $D^-$, to find
\begin{equation}
\label{MUL}
D^+(\e^{-\mi\varphi}\lambda,{\vec v}^{[-1]})
/ D^+(\e^{+\mi\varphi}\lambda,{\vec v}^{[1]})
\vert_{\lambda=-E_{2k}} = -\e^{\mi\varphi/2}.
\end{equation}
Whereas the procedure for $V(q)=q^N$áclosed upon itself at once, 
this general one invokes the complex-rotated potentials $V^{[\pm 1]}$ and,
step by step, all $L$ distinct complex potentials $V^{[\ell]}$ as well,
their (complex) spectra $\{E_{2k}^{[\ell]}\}$ constituting new unknowns 
(they are just analytical continuations of the initial spectrum in complex
${\vec v}$-space, but we cannot exploit this feature constructively). 

Next, the absolute phases in eq.(\ref{MUL}) 
(and its cyclic permutations on the ${\vec v}^{[\ell]}$) are determined 
by reference to the homogeneous case (see \cite{V3}).
This operation is essential to get Bohr--Sommerfeld-lookalike formulae, 
governed by an explicit quantum number $k=0,1,2,\ldots$. 
Our final result, once both parities have been merged again, is then
\begin{eqnarray}
\label{EQC}
\log D^\pm(-\e^{-\mi\varphi}E_{k}^{[\ell]},{\vec v}^{[\ell-1]})
-\log D^\pm(-\e^{+\mi\varphi}E_{k}^{[\ell]},{\vec v}^{[\ell+1]}) =
\qquad \qquad \\
\qquad \qquad = \pi\mi\Bigl[ k+{1 \over 2} \pm {N-2 \over 2(N+2)} \Bigr]\quad 
{\rm for\ } k\ {\scriptstyle{\scriptstyle{\rm even}\atop \scriptstyle{\rm odd}}}
,\ \ell \ {\rm integer\ mod\, }L \nonumber
\end{eqnarray}
the branch of $\log D(\lambda)$ being defined by continuity from 
$(\lambda=0,\ {\vec v}=0)$.

In each parity sector independently,
eqs.(\ref{EQC}) form a system of constraints 
tying each $E_{k}^{[\ell]}$ at fixed $\ell,k$ 
to the two other spectra 
$\{E_{j}^{[\ell-1]}\}$, $\{E_{j}^{[\ell+1]}\}$ 
(whose zeta-regularized products build the left-hand-side determinants). 
The system of all such interacting points is better displayed,
in proper relative positions if each spectrum is first suitably rotated
(we then call it a `chain'): 
the $\ell^{\rm th}$ such chain is then the set
$  \{\e^{\mi\ell\varphi} E_{k}^{[\ell]}\}$, $\ell=0,\ldots,L-1$ (mod $L$)
($k$ being either even or odd, depending on the sector).
Each point of the $\ell^{\rm th}$ chain interacts with every point of
the two adjacent ($(\ell \pm 1)^{\rm th}$) chains, through the logarithm
of their difference entering eq.(\ref{EQC}).

The $\ell^{\rm th}$ equation is now a complex one whenever its unknowns 
$E_{k}^{[\ell]}$ are themselves complex, i.e., for $\ell \ne 0$ or $L/2$:
eq.(\ref{EQC}) then remains a formally `complete' system of mutual constraints
for the unknowns $E_{k}^{[\ell]}$.
As in the homogeneous case we then surmise that eqs.(\ref{EQC}), 
which are exact, also constitute genuine quantization conditions, 
i.e., they have the capacity to specify all their unknowns 
provided the asymptotic condition (\ref{BS}) is also enforced 
upon each spectrum $\{E_{k}^{[\ell]}\}$ separately 
(using the rotated ${\vec v}^{[\ell]}$). 
Whenever this property can be confirmed, 
eqs.(\ref{EQC}) will then determine the spectral problem analytically.

We will now report numerical experiments upon quartic potentials 
which numerically confirm this conjecture in some regions of parameter space,
by achieving {\sl effective\/} computations of the spectra out of eqs.(\ref{EQC}). 

(To concretely apply eqs.(\ref{EQC}), we evaluate the functions 
$\log D^\pm$ through the formula (\ref{McL}) 
(adding higher correction terms to accelerate convergence), and we use Newton's
method to solve eq.(\ref{EQC}) for each $E_{k}^{[\ell]}$ in turn,
but only up to some finite $k_{\rm max}$ beyond which levels are 
semiclassically assigned once for all; 
the height of the cutoff $k_{\rm max}$ regulates the final accuracy.)

\subsection{Even quartic oscillator}

We illustrate the preceding formalism through the quantization of the levels 
for the even potential $V(q)=q^4+v_2 q^2$ on the real line. Parity implies:
that the spectrum splits into even and odd sectors, corresponding 
to the Neumann and Dirichlet problem on the half-line, respectively;
and that the order of symmetry is $L=3$ (instead of 6 generically).

To the initially real $v_2$ giving the real spectrum $\{E_k\}$ 
are then associated:
a complex spectrum $\{E'_k\}$ for the coupling constant $\j v_2$,
and $\{E''_k\}$ for $\j^2 v_2$ (where $\j=\e^{2\pi\mi/3}$). 
The three chains $\{E_k\},\ \j\{E'_k\},\ \j^2\{E''_k\}$ are shown on fig.1
for some values of the harmonic coupling constant $v_2$.
By reality symmetry, there are only two independent 
equations/unknowns (one real, and one complex).

The semiclassical information (\ref{BS}) was input up to 6 terms (O($E^{-1/2}$))
to make the accuracy grow faster with respect to the truncation $k_{\rm max}$.

The numerical procedure is to solve eqs.(\ref{EQC}) in some cyclic sequence 
and manner that will achieve a contractive-looking iteration, 
as in the homogeneous case; there are now many more implementation details
left open, and we tried several.

The homogeneous case ($v_2=0$) enjoyed full ternary rotation symmetry,
and also yielded strong contraction ratios ($\losim 0.4$) \cite{V2,V3}.
The main change seen when $v_2 \ne 0$ is that certain iteration schemes
excite new chain fluctuation modes that are quite asymmetrical
(even about the fully symmetrical $v_2=0$ chain) and much less stable (contraction ratios $\approx \pm 0.9$). 
This holds for the naive (and most symmetrical) deformation 
of the $v_2=0$ iteration, which at each step addresses each chain in turn 
but refreshes all of their values only after a complete $\ell$-cycle. 
Those `bad' modes ruin the quality of convergence already very close to
$v_2=0$.

By contrast, we empirically found that an iteration scheme alternating between
the real and one complex chain and immediately refreshing
their values at each step, remains fairly stable near $v_2=0$
while appearing to converge towards the correct spectra. 
It nevertheless deteriorates as $v_2$ grows, becoming
totally erratic above $v_2 \approx +2$; hence it cannot be carried towards the
harmonic limit. By contrast, the scheme looks much more robust in the
negative direction (double-well region), showing no perceptible degradation
down to  $v_2 \approx -10$~! Its contraction ratios are plotted on fig.2, left
(for $v_2 < 2$ they were estimated from the iteration results and by 
the diagonalization of the linearized dynamics near the fixed point, 
and for $v_2 \ge 2$ only by the latter method).

Some explanation of the results comes from fig.1: when $v_2$ grows, 
the complex chain and its conjugate become almost degenerate towards low
quantum numbers (a sort of complex tunneling effect), increasing the logarithmic pair interactions and the instabilities (the linearized-dynamics
matrix entries grow even faster, like [mutual distance]$^{-1}$).
An obvious improvement is then to decouple the interaction 
between any two complex-conjugate chain points by enforcing this symmetry
as an explicit constraint (immediate refreshing also has to be kept);
this achieves a much more uniform linear stability indeed (fig.2, right).
However, global instabilities still arise, albeit more slowly, 
and we still lost all convergence beyond $v_2 \approx +5$
(those instabilities primarily hit the Newton root-finding algorithm, 
so there is room left for remedies).

The preceding numbers referred to iterations in the even-parity sector.
As fig.2 shows, the odd sector tends to behave more stably.
A conceivably more robust scheme might then be to confine iterations 
to the odd spectrum, then to solve for the even determinants 
from the coupling relations (\ref{DW}) instead.
 
\subsection{Extension to non-even potentials}

In \cite{V4} we showed that the exact quantization formalism was fully valid 
for homogeneous potentials of any odd degree $N$, and specially for $N=1$ 
which has Airy functions both as solutions and as spectral determinants. 
The paradox in the very regular behavior of this case is that 
the underlying potential $V(q)=|q|$ is not even once differentiable at $q=0$.
This exemplifies that the `even' and `odd' spectrum distinguished 
by the exact quantization mechanism are just the Neumann and Dirichlet spectra.
Parity properties of the polynomial defining $V$ are then irrelevant: 
the underlying even potential can and will be $V(|q|)$ in all cases.

However, a generic non-even polynomial $V(q)$ has $Z(0) \ne 0$, 
a case which our approach still handles incorrectly. 
To test the exact quantization conditions (\ref{EQC}) 
upon quartic non-even polynomials, we therefore tried a few with $Z(0) = 0$,
using
\begin{equation}
\label{Z0}
Z(0)  \Bigl [ = (2\pi)^{-1} {\tilde b}_0 \Bigr ] = - {v_3 \over 4}+{v_1 v_2 \over 8}-{v_1^3 \over 32} 
\quad  {\rm for\ } N=4;
\end{equation}
for instance, $V(q) = q^4 - q^3 + |q| /8$ yielded results comparable 
to the previous ones for even systems. 
A further validation is also brought by the next computation.

\section{Exact wave-function resolution}

That extension of the exact quantization formalism beyond even potentials
is important for the following application. 
If no parity property is imposed upon the polynomials $V(q)$ 
with respect to the endpoint $q=0$,
then this endpoint can be taken to an arbitrary value $a$ 
(on the real line, at least), 
thus restoring translational invariance effectively. 
Equivalently, 0 is kept as endpoint but the potential gets shifted to 
$V_a (q) \defi V(q+a)-V(a)$ on the half-line (becoming $V(|q|+a)-V(a)$ on the whole line). 
It further happens, in the special case $ V(q) = q^4 + v_2 q^2 $, that our present restriction 
$Z(0) \equiv 0$ remains fulfilled by all the potentials $V_a (q)$ 
(simply using eq.(\ref{Z0})).
Hence this provides another opportunity to try the current formalism.

So, we now use eq.(\ref{ID}) at $q=a$ instead of 0 and from right to left: 
then it states that the naturally normalized solution of eq.(\ref{Schr}) 
at $q=a$, resp. its first derivative, equal
\begin{equation}
\label{SOL}
\psi(a) \equiv D^-(V(a)+\lambda), \qquad \psi'(a) \equiv -D^+(V(a)+\lambda)
\qquad {\rm under\ the\ potential\ } V_a (q)\ {\rm on\ } [0,+\infty).
\end{equation}

In turn, these determinants are canonically specified as 
the zeta-regularized products over their chains of zeros.
And these are the eigenvalues of $V_a ^{[\ell]}(q)$, hence they obey 
exact quantization conditions of the form (\ref{EQC}) and,
hopefully, this makes them reachable by some convergent iteration scheme.
In this sense, eq.(\ref{SOL}) will provide an algorithmic specification 
to solve eq.(\ref{Schr}) (for arbitrary $\lambda$).

We illustrate this program with a calculation of the ground state eigenfunction
for the homogeneous case $V(q)=q^4$. 
The shifted potential is then $ V_a(q) = q^4 + 4 a q^3 + 6 a^2 q^2 + 4 a^3 q $,
and $\psi(a)$ is the value of its determinant $D^-$ at the point $a^4-E_0$,
where the eigenvalue $E_0 \approx 1.06036209$ is part of the input here
(but the levels themselves are the output of a similar exact calculation
using $V(q)=q^4$).

The exact quantization conditions for $V_a(q)$ now involve 
$L=6$ distinct chains (except at $a=0$, where $L=3$ by parity symmetry): 
$\ell=0,\ 3$ are real, $(1,5),\ (2,4)$ being doublets of complex conjugates. 
We only tested $\psi(a)$, for a few values of $a \ge 0$.
We found this iteration sequence: $\ell=\{0,2,3,1\}$
(then cyclically continued, with immediate refreshment of each chain and
straight enforcement of 1--5 and 2--4 symmetry) to converge indeed for 
$a \losim 1.7$ (contraction ratio per cycle $ \losim 0.67$, 
better than other sequences for no clear reason).
Just as above, instabilities suppressed convergence for higher $a$.
Our output points are plotted on fig.3, against the curve produced by 
a standard integration routine and upon which
only the global normalization was fitted: 
the results show an overall 4--5-digit agreement.

\subsection{Concluding remarks}

We have reduced the resolution of polynomial 
Schr\"odinger equations (\ref{Schr}) (for their eigen/values/vectors) 
to that of a discrete system of selfconsistent exact quantization conditions,
eq.(\ref{EQC}), having as unknowns $(N+2)$ countable chains of points 
subject to asymptotic boundary conditions 
supplied by conventional Bohr--Sommerfeld formulae (\ref{BS})
(still provided 
the zeta function $Z(s)$ of eq.(\ref{Z}) vanishes at the origin.)

We are also getting {\sl numerical\/} evidence that the so reduced problem 
is effectively solvable in some regions of parameter space
through iterations which seem to converge geometrically.
This suggests that the solution is being described as a fixed point for
some explicit action of eqs.(\ref{EQC}) upon infinite but asymptotically
constrained chains, appearing to be contractive
and also to admit robust finite-dimensional approximations.

On the other hand, since eqs.(\ref{EQC}) are very tied to the values of the
degree $N$ and symmetry order $L$, they may be ill-suited to transitional
regions where these numbers jump 
(e.g, $v_2 \to\ 0\ {\rm or\ }+\infty$ in $V(q)=q^4 + v_2 q^2$).
Moreover, our available evidence is currently confined 
to even quartic potentials close to $q^4$ and (from earlier work \cite{V3}) 
to purely homogeneous potentials for their spectrum only 
(but up to quite high degrees). Extensions to higher $N$, $Z(0) \ne 0$, 
arbitrary complex $q$ and $\vec v$ etc., and to more general differential systems, 
are all conceivable but they remain open issues.

Finally we argue that, while the integration of eq.(\ref{Schr}) 
by quadratures alone may still be a remote goal,
another relevant question is how much the set of admissible integration methods needs to be enlarged to reach that purpose. We are then finding that
zeta-regularized infinite products (of order $<1$) 
plus the solving of one type of fixed-point equations 
(which seem to have contractive and other nice properties) appear to be
well adapted, and sufficient to the task already in a few cases.

\medskip

{\bf Acknowledgment:} we are grateful to R. Guida (Saclay) for putting at our
disposal his computer program to calculate hundreds of anharmonic oscillator levels with high accuracy.

\vfill\eject

\centerline{\bf Figure captions}
\bigskip

Fig.~1. Interacting chains
in their equilibrium  positions yielding the exact odd spectra of $V^{[\ell]}$
(points marked with the corresponding $\ell$-value),
for potentials $V(q) = q^4 + v_2 q^2$.
As $v_2 \to -\infty$, the complex chains tend to shadow 
the resonance spectrum of the potential $-|v_2| q^2$;
as $v_2 \to +\infty$, the complex chains tend to shadow
what would be the $\ell=1$ chain for the potential $v_2 q^2$.

\medskip

Fig.~2. Contraction ratios of iteration schemes converging to the spectra of 
$V(q) = q^4 + v_2 q^2$ ($+$: even spectrum; $\circ$: odd spectrum).
Left: numerical estimates for simple iteration scheme 
with immediate refreshing;
we note that the numerical values never reach exactly unity, 
even though no definite conclusion can be inferred 
(no error estimate at all is implied in our data,
but 2--3 digit stability is typical).
Right: numerically estimated {\sl moduli\/} of same ratios but for
an iteration scheme further decoupling conjugate pairs (here, different parts
of the curves may correspond to different branches of linearized eigenvalues, 
some being negative or in complex pairs). 

\medskip

Fig.~3. Naturally normalized ground-state eigenfunction
for the homogeneous quartic potential $q^4$. 
Curve: 
computer integration of the Schr\"odinger equation using the NAG routine D02KEF;
its rescaling factor was the only number fixed by a fit.
$+$: points obtained by iterative exact quantization; 
$\diamond$: numerical estimates for the corresponding contraction ratios. 

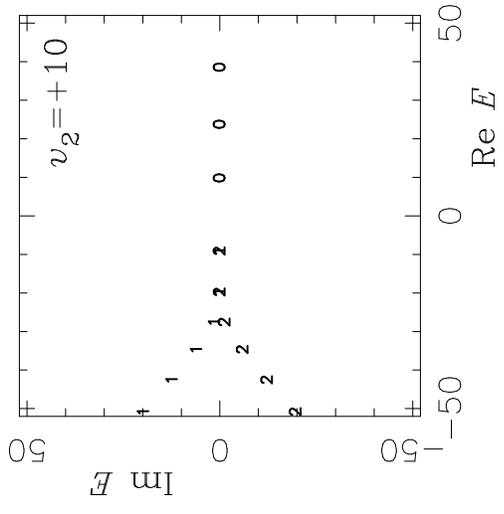
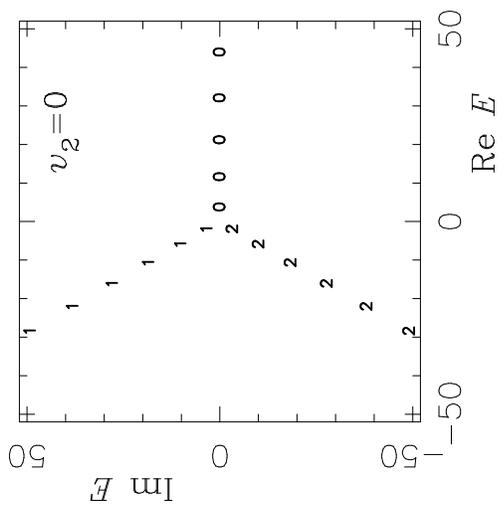
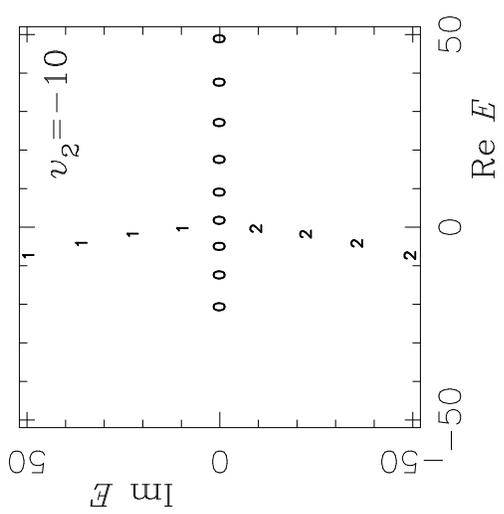
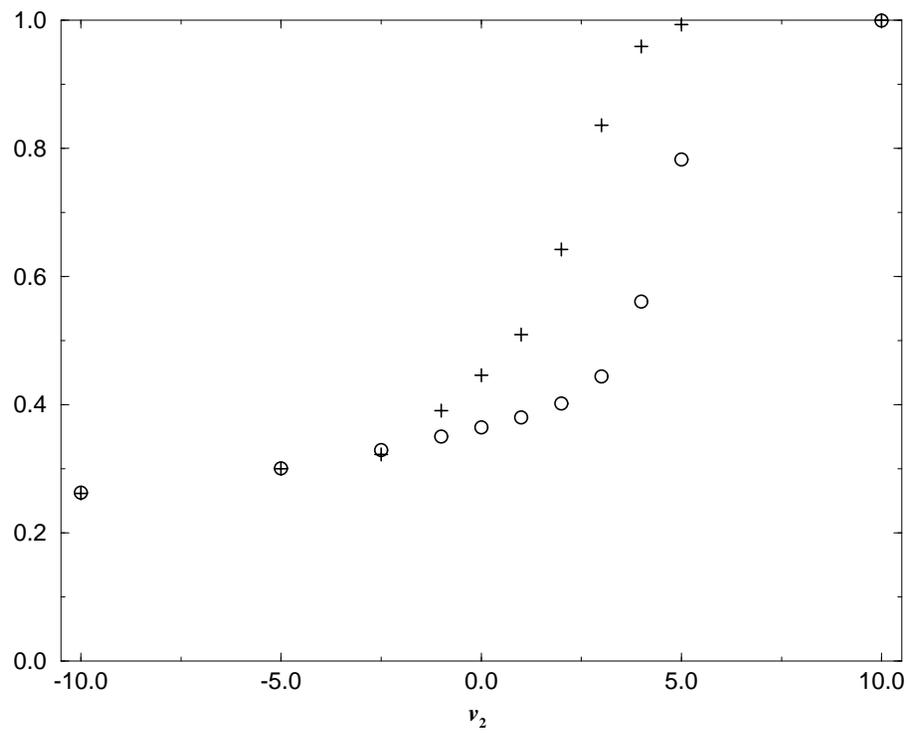
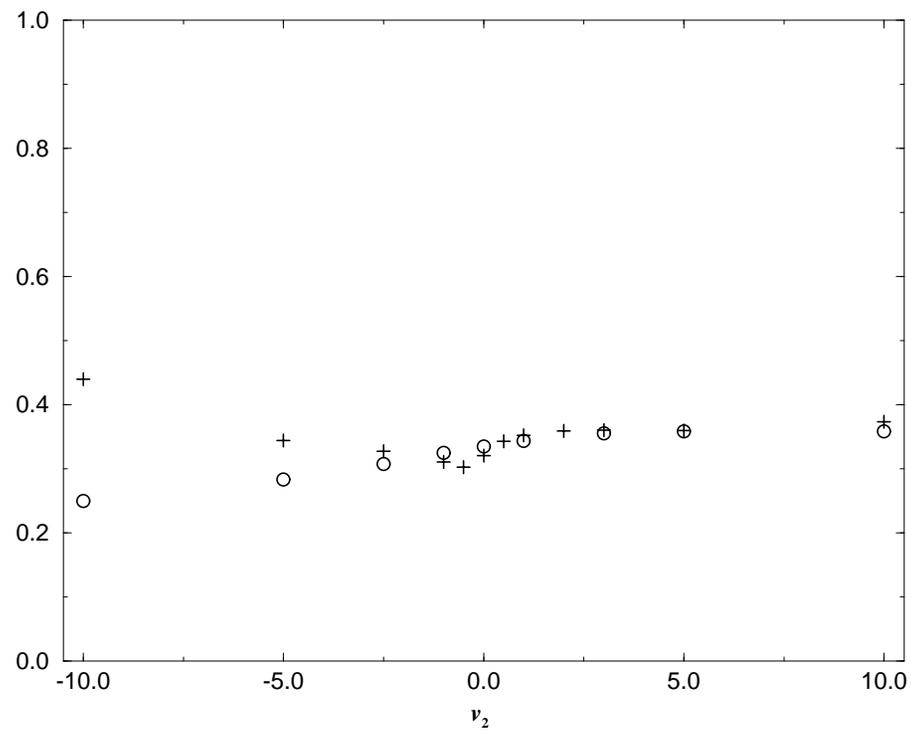
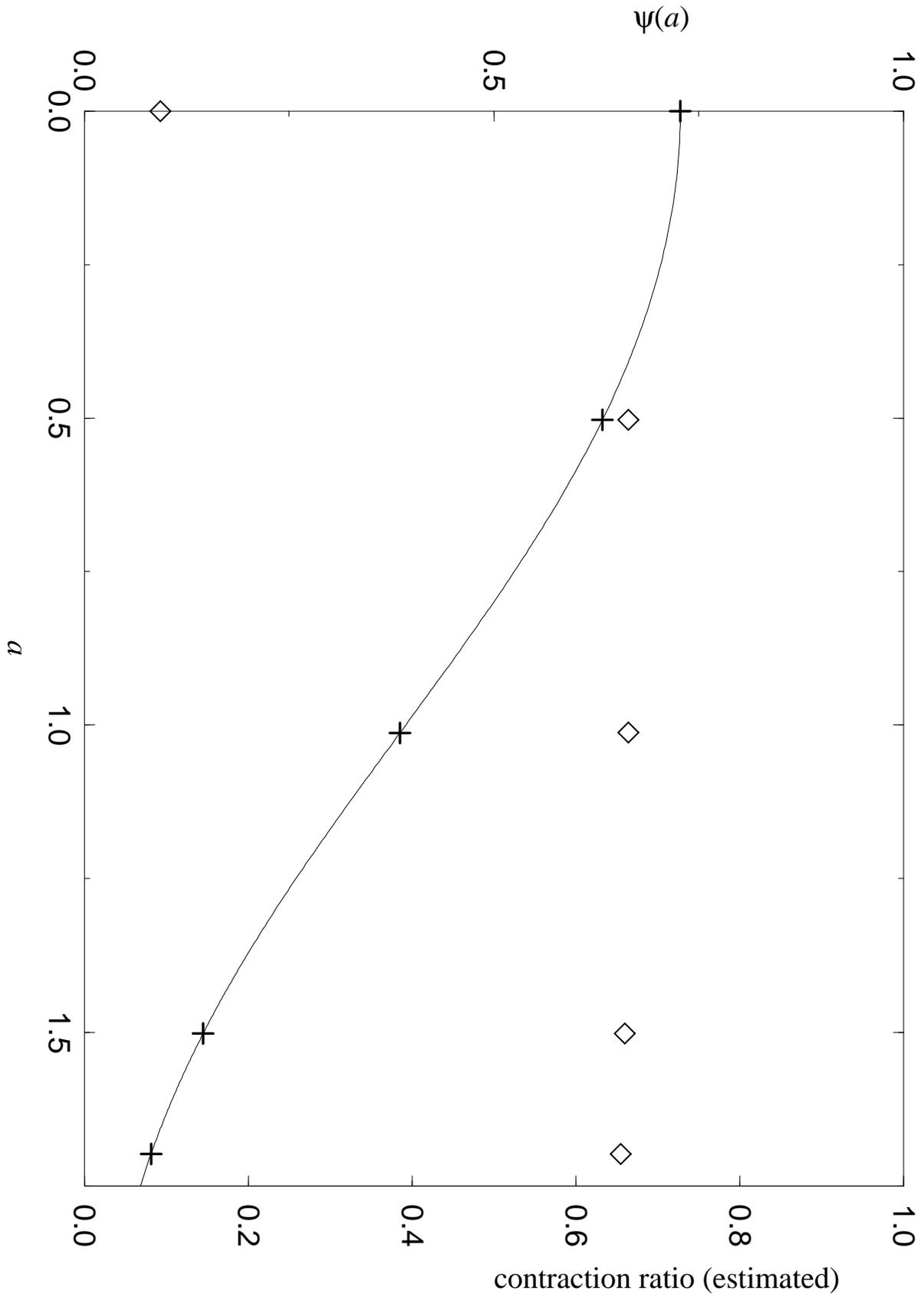
\end{document}